\documentclass[aps,prd,10pt,a4paper,twocolumn,preprintnumbers,nofootinbib,floatfix]{revtex4-1}
\pdfoutput=1
\usepackage[a4paper, hdivide={1.91cm,,1.165cm}, vdivide={1.83cm,,3.4cm}]{geometry}

\usepackage{amstext,amssymb}
\usepackage[intlimits]{amsmath}

\usepackage[ansinew]{inputenc}
\usepackage{graphicx}
\usepackage{units}
\usepackage[hyperfootnotes=false]{hyperref}

\newcommand{\hc}{\ensuremath{\text{h.c.}}}

\allowdisplaybreaks

\begin{document}


\preprint{UCI-TR-2019-21, \href{https://doi.org/10.1103/PhysRevD.100.075027}{Phys. Rev. \textbf{D100} (2019) 075027}, \href{http://arxiv.org/abs/1908.03286}{arXiv:1908.03286}}

\title{Observing Dirac neutrinos in the cosmic microwave background}

\author{Kevork N. Abazajian}
\email{kevork@uci.edu}
\affiliation{Department of Physics and Astronomy, University of California, Irvine, California 92697-4575, USA}

\author{Julian Heeck}
\email{Julian.Heeck@uci.edu}
\affiliation{Department of Physics and Astronomy, University of California, Irvine, California 92697-4575, USA}

\hypersetup{
pdftitle={Observing Dirac neutrinos in the cosmic microwave background},   
pdfauthor={Kevork N. Abazajian, Julian Heeck}
}


\begin{abstract}
Planned CMB Stage IV experiments have the potential to measure the effective number of relativistic degrees of freedom in the early Universe, $N_\text{eff}$, with percent-level accuracy. This probes new thermalized light particles and also constrains possible new-physics interactions of Dirac neutrinos. Many Dirac-neutrino models that aim to address the Dirac stability, the smallness of neutrino masses or the matter--anti-matter asymmetry of our Universe endow the right-handed chirality partners $\nu_R$ with additional interactions that can thermalize them. Unless the reheating temperature of our Universe was low, this leads to testable deviations in $N_\text{eff}$. We discuss well-motivated models for $\nu_R$ interactions such as gauged $U(1)_{B-L}$ and the neutrinophilic two-Higgs-doublet model, and compare the sensitivity of SPT-3G, Simons Observatory, and CMB-S4 to other experiments, in particular the LHC.
\end{abstract}

\maketitle

\section{Introduction}

The sensitivity of anisotropies in the cosmic microwave background (CMB) to extra radiation density like that in the form of effective extra numbers of neutrinos $N_\text{eff}$ has been known for some time~\cite{Jungman:1995bz}. Upcoming limits from the CMB and large-scale structure on extra radiation from the early Universe are entering a qualitatively new regime, with sensitivity to particle species that have decoupled from equilibrium at very early times and high energy scales. In this article, we show that there are direct implications of this sensitivity to neutrino mass models. 

Any extra non-photon radiation energy density $\rho_\text{rad}$ is usually normalized to the number density of one active neutrino flavor, $N_\text{eff} \equiv (8/7) \left(11/4\right)^{4/3}\rho_\text{rad}/\rho_\gamma$. The current Planck measurement is $N_\text{eff}=2.99\pm 0.17$ (including baryon acoustic oscillation (BAO) data)~\cite{Aghanim:2018eyx}, perfectly consistent with the Standard Model (SM) expectation $N_\text{eff}^\text{SM} = 3.045$~\cite{Mangano:2005cc,Grohs:2015tfy,deSalas:2016ztq}.
CMB Stage IV (CMB-S4) experiments have the potential to constrain $\Delta N_\text{eff} \equiv N_\text{eff}-N_\text{eff}^\text{SM} = 0.060$ (at 95\% C.L.)~\cite{Abazajian:2016hbv,Abazajian:2019eic}, which is very sensitive to new \emph{light} degrees of freedom that were in equilibrium with the SM at some point, even if it decoupled at multi-TeV temperatures. Indeed, a relativistic particle $\phi$ that decouples from the SM plasma at temperature $T_\text{dec}$ contributes
\begin{align}
\Delta N_\text{eff} &\simeq 0.027 \left(\frac{106.75}{g_\star (T_\text{dec})}\right)^{4/3}  g_s\,,
\label{eq:DeltaNeff}
\end{align}
where $g_s$ is the number of spin degrees of freedom of $\phi$ (multiplied by $7/8$ for fermions) and $g_\star (T_\text{dec})$ is the sum of all relativistic degrees of freedom except $\phi$ at $T=T_\text{dec}$. At temperatures above the electroweak scale, $g_\star$ saturates to $106.75$, the maximum amount of entropy available from SM particles.
Reference~\cite{Baumann:2016wac} has recently studied the impact of CMB-S4 on axions and axion-like particles ($g_s=1$), which are reasonably well motivated but could easily lead to an entropy-suppressed contribution $\Delta N_\text{eff}\simeq 0.027$ that is below the CMB-S4 reach.

It should be kept in mind, however, that an even better motivation for light degrees of freedom comes from the discovery of non-zero neutrino masses:  if neutrinos are \emph{Dirac} particles then we necessarily need two or three effectively massless chirality partners $\nu_R$ in our world, which would contribute a whooping $\Delta N_\text{eff}\geq  2\times 0.047=0.09$ (two $\nu_R$) or even $\Delta N_\text{eff}\geq 0.14$ (three $\nu_R$) if thermalized with the SM, easily falsifiable or detectable! 
While it is well known that just SM + Dirac $\nu$ does not put $\nu_R$ in equilibrium due to the tiny Yukawa couplings $m_\nu/\langle H \rangle\lesssim 10^{-11}$~\cite{Shapiro:1980rr,Antonelli:1981eg}, one often expects additional interactions for $\nu_R$ in order to explain the smallness of neutrino masses, to generate the observed matter--anti-matter asymmetry of our Universe, and to protect the Dirac nature from quantum gravity, as we will highlight below.
All of these new $\nu_R$ interactions will then face strong constraints from CMB-S4 that will make it difficult to see the mediator particles in any other experiment, in particular at the LHC.

The basic idea to measure new interactions via $N_\text{eff}$ in Big Bang nucleosynthesis (BBN) or the CMB is of course old~\cite{Steigman:1979xp,Olive:1980wz}, see for example the reviews~\cite{Dolgov:1981hv,Olive:1999ij,Dolgov:2002wy}. It is timely to revisit these limits though since we are on the verge of reaching an important milestone: sensitivity to Dirac-neutrino induced $\Delta N_\text{eff}$ even if the $\nu_R$ decoupled above the \emph{electroweak} phase transition!
As we will outline in this article, the non-observation of any $\Delta N_\text{eff}$ in CMB-S4 will then have serious consequences for almost all Dirac-neutrino models, in particular those addressing the origin of the small neutrino mass.

The rest of this article is organized as follows: Sec.~\ref{sec:experiments} gives a brief overview of the current measurements of $N_\text{eff}$ and future reach. In Sec.~\ref{sec:theory} we discuss the impact of stronger $\Delta N_\text{eff}$ limits on a number of Dirac-neutrino mass models. We conclude in Sec.~\ref{sec:conclusion}.

\section{Observing \texorpdfstring{$N_\text{eff}$}{Neff}}
\label{sec:experiments}

The CMB is sensitive to the radiation energy density of the Universe via the variant effects of radiation on the features of the acoustic peaks of the CMB and its damping tail. The acoustic scale of the CMB is altered inversely proportionally to the Hubble rate at the time of last scattering, $\theta_\text{sound} \propto H^{-1}$, while the scattering causing the exponentially suppressed damping tail of the CMB anisotropies goes as $\theta_\text{damping} \propto H^{-1/2}$. These differential effects provide the primary signatures of extra $\Delta N_\text{eff}$ in the CMB power spectrum. The primordial helium abundance, $Y_p$, also changes the scales of  $\theta_\text{sound}$ to $\theta_\text{damping}$ similarly, however the near degeneracy between $N_\text{eff}$ and $Y_p$ is broken by other physical effects, including the early integrated Sachs--Wolfe effect, effects of a high baryon fraction, as well as the acoustic phase shift of the acoustic oscillations \cite{Hou:2011ec,Follin:2015hya}. 

The limit from Planck plus BAO data is $N_\text{eff}=2.99\pm 0.17$~\cite{Aghanim:2018eyx}, where the limit is from a single parameter extension of the standard $\Lambda$CDM 6-parameter cosmological model. We translate this into a $2\sigma$ constraint $\Delta N_\text{eff} < 0.28$.
Currently underway and future experiments are forecast to have even greater sensitivity, even with more conservative assumptions about the possible presence of new physics. The South Pole Telescope SPT-3G is a ground-based telescope currently in operation, with a factor of $\sim 20$ improvement over its predecessor. SPT-3G is forecast to have a sensitivity of $\sigma(\Delta N_\mathrm{eff}) = 0.058$, given here as the single standard deviation ($1 \sigma$) sensitivity \cite{Benson:2014qhw}. This sensitivity is conservative in that it includes the variation of a nine-parameter model for all of the new physics which SPT-3G will be tackling: $\Lambda$CDM (six parameters), $N_\text{eff}$, active neutrino mass ($\Sigma m_\nu$), plus tensors. 
We estimate the $2\sigma$ sensitivity of SPT-3G as $\Delta N_\text{eff} < 0.12$. The CMB Simons Observatory (SO), which will see first light in 2021, is forecast to have $1\sigma$ sensitivity in the range of $\sigma(\Delta N_\mathrm{eff}) = 0.05\ \text{to}\ 0.07$~\cite{Abitbol:2019nhf}.

For the noise level and resolution of CMB-S4, the differential effects on the acoustic peaks and damping tail are predominately measured through the $TE$ spectrum at multipoles $\ell > 2500$~\cite{Abazajian:2019eic}. The sensitivity of CMB-S4 is forecast to be $\Delta N_\text{eff} = 0.060$ at 95\% C.L., as a single parameter extension to $\Lambda$CDM.

\begin{figure}[tb]
\includegraphics[width=0.49\textwidth]{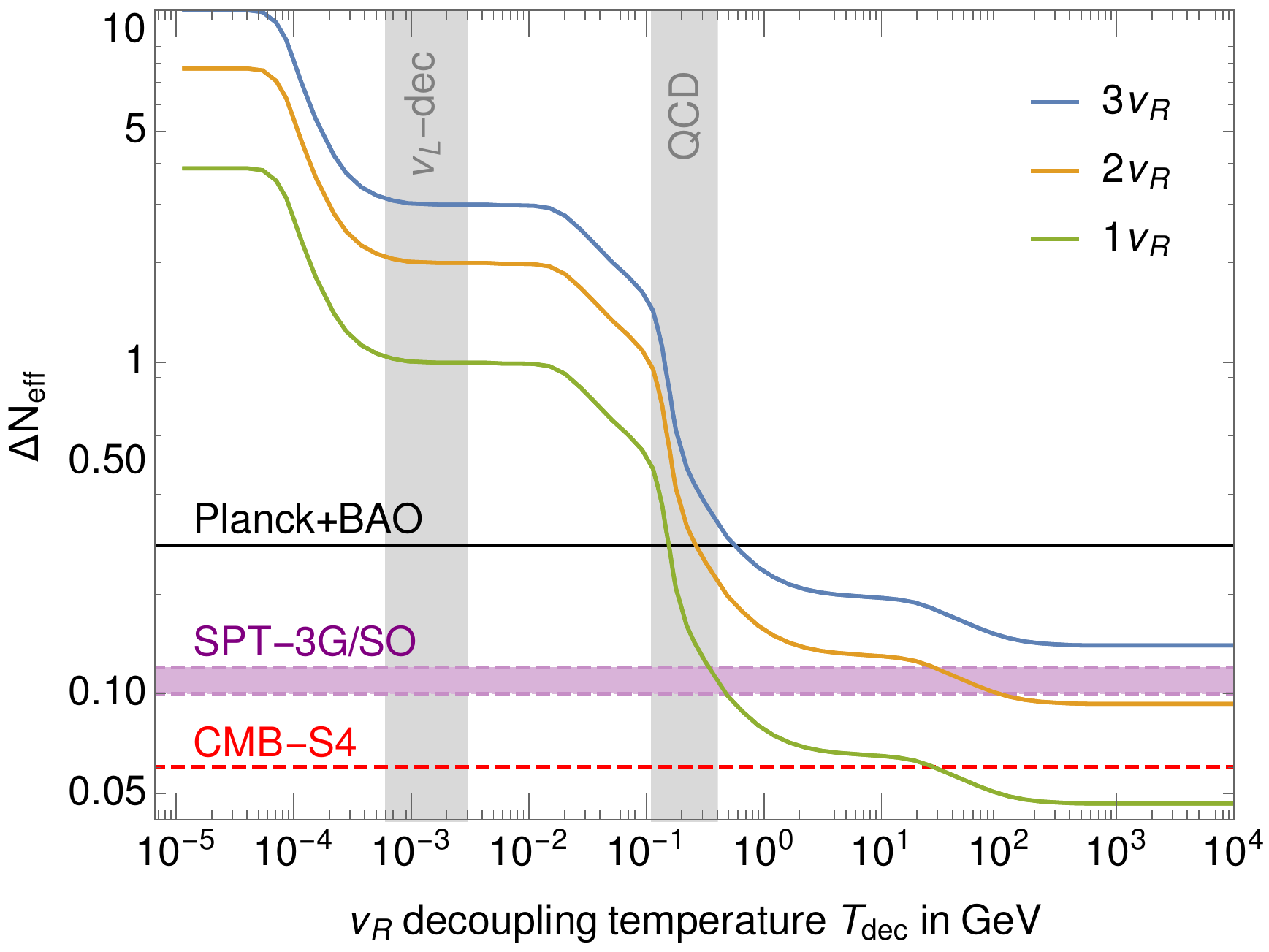}
\caption{Contribution of one, two or three right-handed neutrinos $\nu_R$ to $\Delta N_\text{eff}$ as a function of their common decoupling
temperature $T_\text{dec}$.
The horizontal lines indicate the current $2\sigma$ limit from Planck+BAO as well as the projected reach of SPT-3G, SO, and CMB-S4.	
}
\label{fig:deltaNeff}
\end{figure}

In Fig.~\ref{fig:deltaNeff} we show the current $2\sigma$ limit on $N_\text{eff}$ as well as the SPT-3G, SO, and CMB-S4 forecast as a function of the decoupling temperature $T_\text{dec}$ using Eq.~\eqref{eq:DeltaNeff}. The current Planck limit requires $T_\text{dec} \gtrsim \unit[0.55]{GeV}$ for three right-handed neutrinos, whereas SPT-3G, SO, and CMB-S4 can conclusively probe this scenario for arbitrary decoupling temperatures!
If only \emph{two} $\nu_R$ are in equilibrium, then SPT-3G/SO can probe  $T_\text{dec} \sim \unit[30]{GeV}$ and CMB-S4 is required to reach arbitrary decoupling temperatures. It is then clear that SPT-3G, SO, and CMB-S4 provide a significant sensitivity to the new physics of Dirac-neutrino models.

\section{Impact on Dirac neutrino models}
\label{sec:theory}

In the following we will discuss the impact a near-future constraint $\Delta N_\text{eff}<0.06$ would have on models involving \emph{Dirac} neutrinos, which automatically bring two to three relativistic states $\nu_R$ that could be in equilibrium and contribute to $N_\text{eff}$. As with all constraints from cosmology, our conclusions rest on additional assumptions regarding the cosmological evolution, namely:
\begin{enumerate}
\item We assume general relativity and the cosmological standard model $\Lambda$CDM.
\item We assume that the (reheating) temperature of the Universe reached at least the mass of the particles that couple to $\nu_R$. This is a strong assumption since we technically only know that the Universe was at least $\sim\unit[5]{MeV}$ hot~\cite{deSalas:2015glj}, everything beyond being speculation. Note however that most solutions to the matter--anti-matter asymmetry require at least electroweak temperatures in order to thermalize sphalerons. Dark matter production also typically requires TeV-scale temperatures, at least for weakly interacting massive particles.
\item No significant entropy dilution. To dilute three $\nu_R$ down to $\Delta N_\text{eff}<0.06$ via Eq.~\eqref{eq:DeltaNeff} one would need to roughly double the SM particle content. This means that Dirac neutrinos would evade $N_\text{eff}$ constraints if they decoupled at temperatures above the hypothetical supersymmetry or grand-unified-theory breaking scales, as both of these SM extensions bring a large number of new particles with them. A different way to generate entropy comes from an early phase of matter domination, which requires a heavy particle that goes out of equilibrium while relativistic and then decays sufficiently late so it has time to dominate the energy density of the Universe~\cite{Scherrer:1984fd,Bezrukov:2009th}.
\end{enumerate}

Note that even if the $\nu_R$ never reached thermal equilibrium, it is possible that they were created non-thermally and still leave an imprint in $N_\text{eff}$~\cite{Chen:2015dka}. Following Refs.~\cite{Chen:2015dka,Zhang:2015wua,Huang:2016qmh} it might even be possible to distinguish this $\nu_R$ origin of $\Delta N_\text{eff}$ by observation of the cosmic neutrino background, e.g.~with PTOLEMY~\cite{Baracchini:2018wwj}. This will not be discussed here.

We will further restrict our discussion to renormalizable UV-complete quantum field theories.
An alternative approach would be to study higher-dimensional operators of an effective field theory with SM fields + Dirac-$\nu$ and put constraints on the Wilson coefficients, e.g.~on the Dirac-$\nu$ magnetic moments~\cite{delAguila:2008ir,Aparici:2009fh,Bhattacharya:2015vja,Liao:2016qyd,Bischer:2019ttk,Alcaide:2019pnf}. However, higher-dimensional operators will give $\nu_R$ production rates that are dominated by the highest available temperature and thus depend explicitly on it~\cite{Baumann:2016wac}. In any renormalizable realization of such operators this growing rate would be cured once the underlying mediators go into equilibrium, which then brings us back to the approach pursued here.

Before moving on to the impact of $N_\text{eff}$ measurements on Dirac neutrino models, let us briefly comment on associated cosmological signatures that arise in our Dirac-neutrino setup. At high temperatures, three $\nu_R$ simply contribute to $N_\text{eff}$ as relativistic particles, as discussed above. However, since they have the same mass as the active neutrinos but a lower temperature, $T_{\nu_R} = (\Delta N_\text{eff}/3)^{1/4} T_{\nu_L} $, they will become non-relativistic slightly \emph{before} the active neutrinos and thus modify the usual neutrino free-streaming behavior by introducing an additional scale. Once the $\nu_L$ also turn non-relativistic we find the total neutrino energy density
\begin{align}
\Omega_\nu h^2 \simeq \left[1+ \left(\frac{\Delta N_\text{eff}}{3}\right)^{3/4}\right]\frac{\sum_{j=1}^3  m_{\nu_j}}{\unit[94]{eV}} \,,
\label{eq:Omega_nu}
\end{align}
which is at least 10\% larger compared to the case of non-thermalized $\nu_R$. Equation~\eqref{eq:Omega_nu} would provide an excellent test of the Dirac-neutrino origin of a measured $\Delta N_\text{eff}$ if the sum of neutrino masses could be determined independently, for example by measuring the absolute neutrino mass scale in KATRIN~\cite{Aker:2019uuj} and the mass hierarchy in oscillation experiments.
The contribution of the $\nu_R$ can be matched to a small effective sterile neutrino mass $m_{\nu,\text{sterile}}^\text{eff} =\left(\Delta N_\text{eff}/3\right)^{3/4} \sum_{j=1}^3  m_{\nu_j}$, as defined and constrained in combination with $N_\text{eff}$ by Planck~\cite{Aghanim:2018eyx}.
As of now, the cosmological neutrino mass measurements obtained via $\Omega_\nu$ are less helpful to constrain Dirac neutrinos than $\Delta N_\text{eff}$, although the increased precision on $\Omega_\nu$ in CMB-S4~\cite{Abazajian:2019eic} and DESI~\cite{Aghamousa:2016zmz} will still provide useful information.

\subsection{\texorpdfstring{$U(1)_{B-L}$}{U(1)B-L} and other gauge bosons}

One important task of Dirac-neutrino model building is to protect the Dirac nature, i.e.~to forbid any and all $\Delta L=2$ Majorana mass terms for the neutrinos. While this can easily be achieved by imposing a global lepton number symmetry $U(1)_L$ on the Lagrangian, there is the looming danger that quantum gravity might break such global symmetries~\cite{Banks:2010zn}. To protect the Dirac nature from quantum gravity it might then be preferable to use a \emph{gauge symmetry} to distinguish neutrino from anti-neutrino. The simplest choice is $U(1)_{B-L}$, which is already anomaly-free upon introduction of the three $\nu_R$ that we need for Dirac neutrino masses. For unbroken $U(1)_{B-L}$ the $Z'$ gauge boson can still have a St\"uckelberg mass, a scenario  discussed in Refs.~\cite{Feldman:2011ms,Heeck:2014zfa}.\footnote{Constraints on a $U(1)_{B-L}$ with \emph{Majorana} neutrinos have been discussed extensively in the literature, e.g.~in Refs.~\cite{Carlson:1986cu,Harnik:2012ni,Ilten:2018crw,Bauer:2018onh}.}

In a more extended scenario one can even break $U(1)_{B-L}$ spontaneously, as long as it is by more than two units in order to forbid Majorana mass terms~\cite{Heeck:2015pia}. The simplest example given in Ref.~\cite{Heeck:2013rpa} has a spontaneous symmetry breaking $U(1)_{B-L}\to \mathbb{Z}_4$, where the remaining discrete gauge symmetry protects the Dirac nature of the neutrinos and the $\Delta (B-L)=4$ interactions allow for leptogenesis~\cite{Heeck:2013vha}, as discussed below.
This \emph{broken} $U(1)_{B-L}$ scenario also allows an embedding into larger gauge groups such as left--right, Pati--Salam or $SO(10)$~\cite{Heeck:2015pia}.

\begin{figure}[t]
	\includegraphics[width=0.49\textwidth]{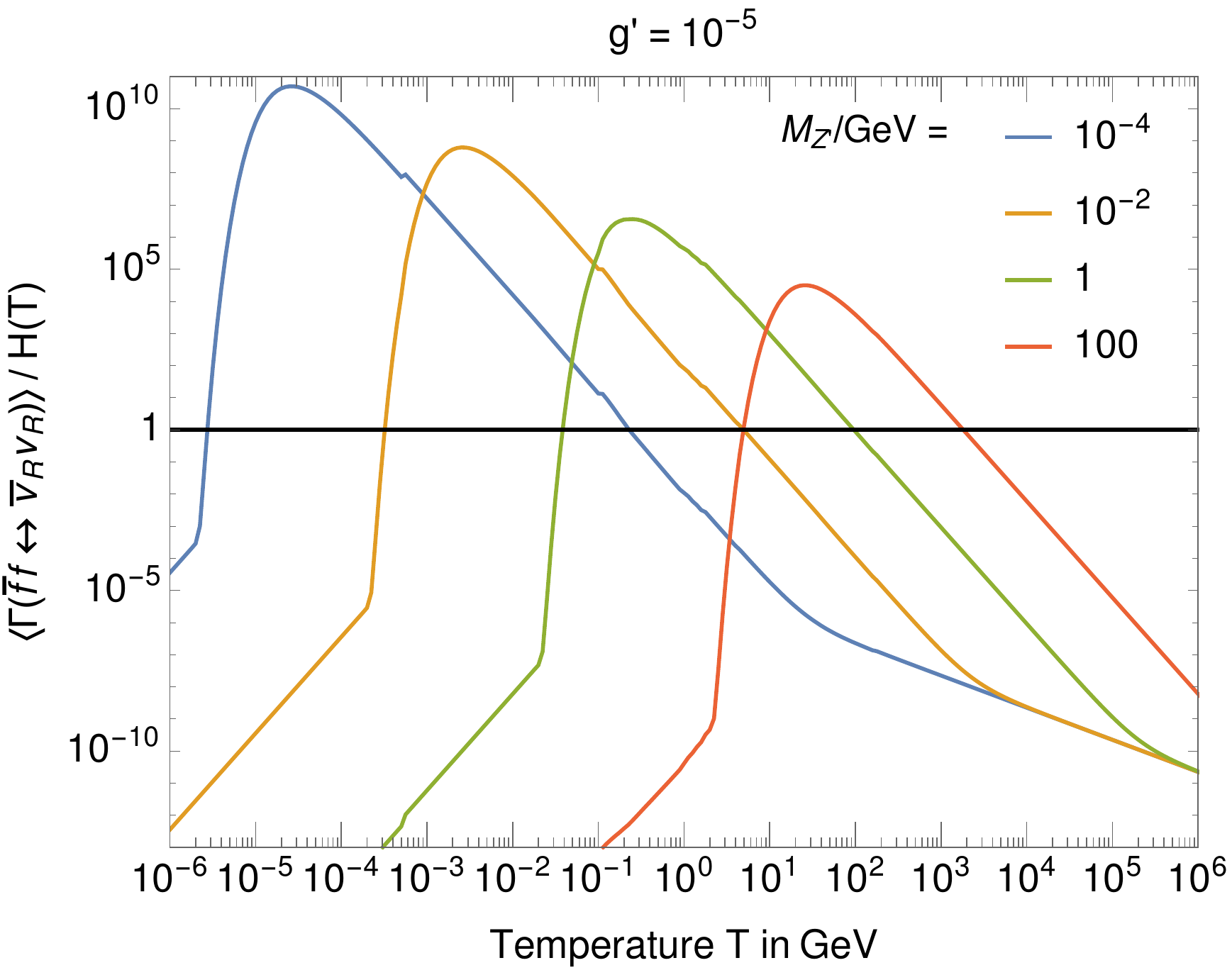}
	\caption{Thermally averaged rate $\langle \Gamma (\bar{f}f\leftrightarrow \bar{\nu}_R\nu_R)\rangle$~\cite{Heeck:2014zfa} divided by the Hubble rate $H(T)$ as a function of temperature for some values of $U(1)_{B-L}$ gauge coupling $g'$ and $Z'$ mass $M_{Z'}$. The $\nu_R$ are thermalized in the region above the horizontal black line.
	}
	\label{fig:production_rate}
\end{figure}

Protecting the Dirac nature of neutrinos in its strongest form thus requires $\nu_R$ couplings to new gauge bosons, the most minimal example being a $Z'$ from $U(1)_{B-L}$. These new gauge bosons can then lead to a thermalization of $\nu_R$ with the rest of the SM plasma in the early Universe, e.g.~via $s$-channel processes $\bar{f} f \leftrightarrow \bar{\nu}_R\nu_R$~\cite{Barger:2003zh,Anchordoqui:2011nh,Anchordoqui:2012qu,SolagurenBeascoa:2012cz}, which then increases $N_\text{eff}$. Equilibrium is attained when this rate $\Gamma$ exceeds the Hubble rate $H(T)\sim T^2/M_\text{Pl}$ at a certain temperature.
The behavior of $\Gamma/H(T)$ is shown in Fig.~\ref{fig:production_rate}, using Eq.~(12) from Ref.~\cite{Heeck:2014zfa}. As can be seen, the ratio $\Gamma/H(T)$ is largest at the temperature $T\sim M_{Z'}/3$, where inverse decays of $Z'$ are highly efficient, so the most aggressive assumption is that the Universe reached this temperature.
Notice that a light $Z'$ will itself start to contribute to $N_\text{eff}$~\cite{Masso:1994ww,Ahlgren:2013wba}.

\begin{figure}[t]
\includegraphics[width=0.49\textwidth]{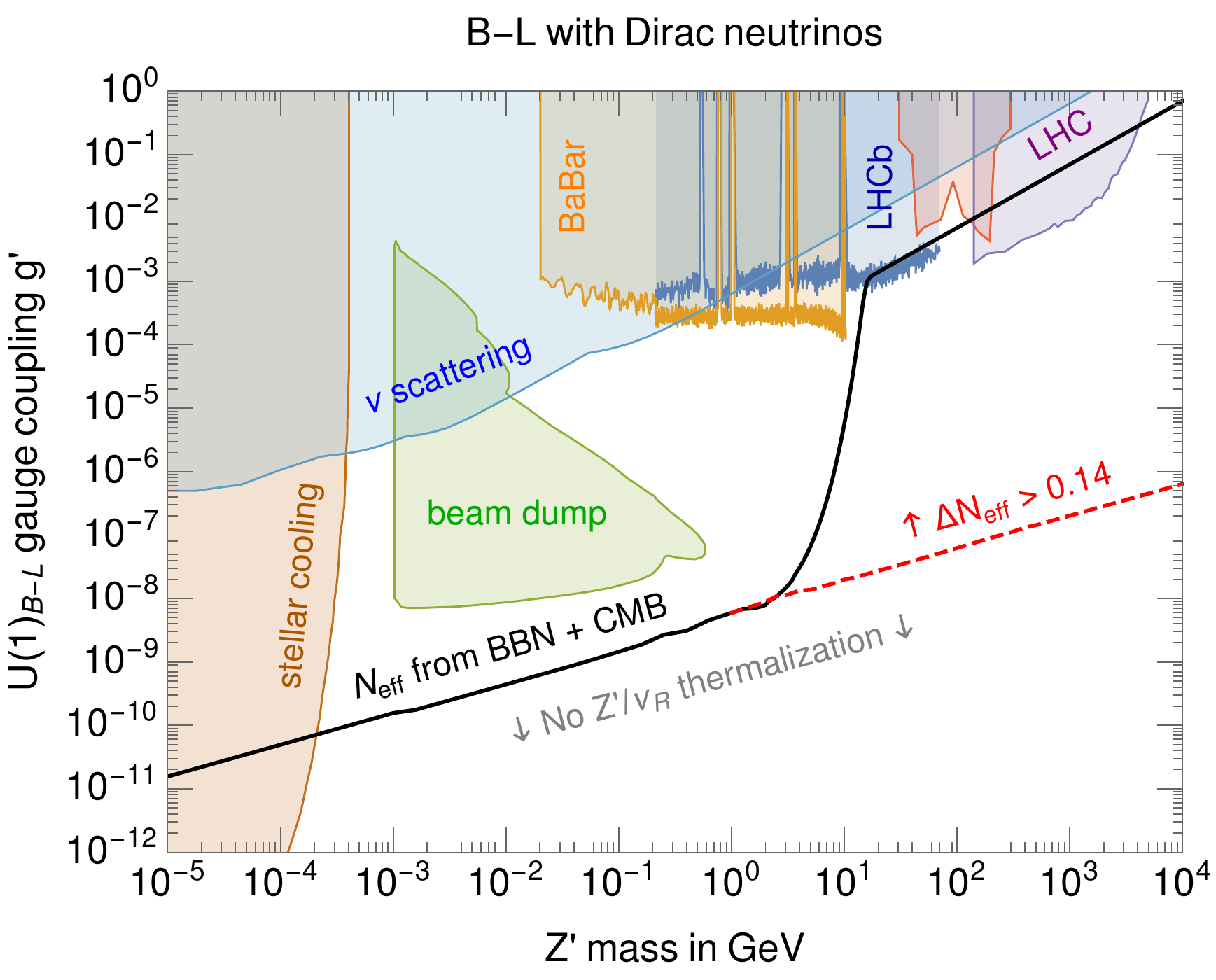}
\caption{Parameter space of a $U(1)_{B-L}$ gauge boson $Z'$ with Dirac neutrinos.
Existing constraints are from stellar cooling~\cite{Redondo:2013lna}, BaBar~\cite{Lees:2014xha}, LHCb~\cite{Aaij:2017rft}, LHC~\cite{Aaboud:2017buh,Escudero:2018fwn}, neutrino scattering~\cite{Bilmis:2015lja,Lindner:2018kjo}, and beam dumps~\cite{Ilten:2018crw,Bjorken:2009mm,Andreas:2012mt,Blumlein:2013cua}.
Not shown are SN1987A constraints that should lie just below the beam-dump region~\cite{Chang:2018rso}.
The BBN+CMB is updated from Ref.~\cite{Heeck:2014zfa}, the red dashed line is the expected reach of SPT-3G, SO, and CMB-S4.
}
\label{fig:B-L_exclusions}
\end{figure}

For heavy $Z'$ masses above $\unit[20]{GeV}$,  we demand that the $\nu_R$ go out of equilibrium before $T\sim\unit[0.5]{GeV}$ (Fig.~\ref{fig:deltaNeff}), which corresponds to the constraint $M_{Z'}/g' > \unit[14]{TeV}$, far better than pre-Planck limits~\cite{Barger:2003zh,Anchordoqui:2011nh,Anchordoqui:2012qu,SolagurenBeascoa:2012cz,Heeck:2014zfa}. A similar limit was recently derived in Ref.~\cite{FileviezPerez:2019cyn}.
For masses $\unit{MeV} < M_{Z'} \lesssim \unit[10]{GeV}$ the limit becomes much stronger due to the $s$-channel resonance of the rate, or equivalently the efficient inverse decay of $Z'$. Here we demand that the $\nu_R$ are out of equilibrium for all temperatures between MeV and $T\sim\unit[0.5]{GeV}$.
For $Z'$ masses below MeV it becomes possible for the $\nu_R$ to go into equilibrium \emph{below} $T\sim\unit{MeV}$, leaving BBN unaffected. However, even in this case the thermalization of $\nu_L$, $\nu_R$, and $Z'$ after $\nu_L$ decoupling would leave an impact on $N_\text{eff}$~\cite{Berlin:2017ftj,Berlin:2018ztp}, already excluded by CMB data. As a result, we have to forbid $\nu_R$/$Z'$ thermalization for all temperatures between eV (CMB formation) and $T\sim\unit[0.5]{GeV}$, which gives the black exclusion line in Fig.~\ref{fig:B-L_exclusions}, updating Ref.~\cite{Heeck:2014zfa}.

This existing $N_\text{eff}$ constraint is stronger than most laboratory experiments, except for dilepton searches at the LHC. 
If future measurements in SPT-3G, SO, and CMB-S4 push the $\Delta N_\text{eff}$ bound below $0.14$, the limits on $Z'$ will change dramatically to
\begin{align}
g' < 2\times 10^{-7}\sqrt{M_{Z'}/\unit{TeV}} \,,
\end{align}
shown as a red dashed line in Fig.~\ref{fig:B-L_exclusions}, because we have to demand that the $\nu_R$ were \emph{never} in equilibrium with the SM. Once again, this limit assumes that the Universe reached a temperature of at least $T\sim M_{Z'}/3$, otherwise the bound weakens. Keeping these assumptions in mind it is clear from Fig.~\ref{fig:B-L_exclusions} that the non-observation of $\Delta N_\text{eff}$ in future CMB experiments will make it impossible to find a $Z'$ coupled to Dirac neutrinos in any laboratory experiment. Turning this around, the observation of a $U(1)_{B-L}$ gauge boson in a collider or scattering experiment would then prove that neutrinos are Majorana particles.

This conclusion is not limited to $B-L$ but extends to other $Z'$~\cite{Barger:2003zh,Chen:2006hn,Anchordoqui:2011nh,Anchordoqui:2012qu,SolagurenBeascoa:2012cz,FileviezPerez:2019cyn} or $W'$~\cite{Steigman:1979xp,Olive:1980wz,Bolton:2019bou} models.
In general, new gauge interactions of $\nu_R$ will face strong constraints from CMB-S4 that will make it difficult to see the gauge bosons, say $Z'$ or $W_R$, in any other experiment, in particular at the LHC.

\subsection{Neutrinophilic 2HDM and other mass models}

Extending the SM by two or three gauge singlets $\nu_R$ allows for Yukawa couplings with the Brout--Englert--Higgs doublet $H$
\begin{align}
\mathcal{L} = y_{\alpha\beta} \overline{L}_\alpha H \nu_{R,\beta} + \hc ,
\label{eq:Dirac_Yukawa}
\end{align}
which give a Dirac-neutrino mass matrix $m_\nu  =  y \langle H\rangle$ after electroweak symmetry breaking. The overall neutrino mass scale is still unknown, but upper bounds between $\unit[0.1]{eV}$ and $\unit[0.2]{eV}$ can be obtained from cosmology~\cite{Aghanim:2018eyx,Vagnozzi:2017ovm,RoyChoudhury:2019hls}, which in turn require us to consider Yukawa couplings $y = m_\nu/\langle H\rangle \lesssim 10^{-12}$. This is a million times smaller than the already-small electron Yukawa and is considered an unappealing fine tuning by most theorists~\cite{Roncadelli:1983ty}. This has spawned a vast literature of models that generate small Dirac masses via other mechanisms and most importantly without the use of small couplings.
The general idea is to forbid the coupling of Eq.~\eqref{eq:Dirac_Yukawa} by means of an additional symmetry~\cite{Roncadelli:1983ty,Davoudiasl:2005ks,Chen:2006hn} and instead couple $\nu_R$ to new mediator particles that eventually also couple to $\nu_L$ and thus create a Dirac mass, often suppressed by loop factors~\cite{Mohapatra:1987nx,Yao:2018ekp} or mediator mass ratios~\cite{Roncadelli:1983ty,Roy:1983be,Ma:2016mwh,CentellesChulia:2018gwr}.

The crucial point is that the new mediator particles unavoidably couple to $\nu_L$ with non-tiny couplings, which thermalizes them in the early Universe at temperatures around their mass. In order to connect $\nu_L$ to $\nu_R$, some of the mediators  also have gauge interactions under $SU(2)_L\times U(1)_Y$.
Since they also have non-tiny couplings with $\nu_R$ by construction, this puts the $\nu_R$ in thermal equilibrium with the SM. Unlike the $Z'$ model of the previous section it makes little sense here to consider couplings that are too small to reach $\nu_R$ equilibrium, as this would defeat the purpose of these models. The only way to evade $N_\text{eff}$ constraints is then to assume that the Universe never reached temperatures of order of the mediator masses.

As an explicit and rather minimal example let us consider the neutrinophilic two-Higgs-doublet model (2HDM)~\cite{Wang:2006jy,Gabriel:2006ns,Sher:2011mx,Zhou:2011rc,Davidson:2009ha}, which introduces a second scalar doublet $\phi$ that exclusively couples to $\nu_R$ by means of a new symmetry:
\begin{align}
\mathcal{L} = \kappa_{\alpha\beta} \overline{L}_\alpha \phi \nu_{R,\beta} + \hc
\end{align}
All \emph{charged} fermions obtain their mass from the main doublet $H$ with vacuum expectation value around $\unit[174]{GeV}$, but the neutrinos obtain a Dirac mass $m_\nu =  \kappa \langle \phi\rangle$. Instead of using small Yukawa couplings it is then possible to simply have a smaller vacuum expectation value for the neutrinos, e.g.~$\langle \phi\rangle \sim\unit{eV}$. The Yukawa couplings $ \kappa$ can then be large, apparently resolving the unwelcome fine tuning of Eq.~\eqref{eq:Dirac_Yukawa}.
Of course, the $ \kappa$ will in particular be large enough to thermalize the $\nu_R$, seeing as $\phi$ is an electroweak doublet that is certainly in equilibrium with the SM at temperatures around $m_\phi$~\cite{Davidson:2009ha}.
The neutrinophilic 2HDM thus predicts $\Delta N_\text{eff} > 0.09$ (two $\nu_R$) or $\Delta N_\text{eff} > 0.14$ (three $\nu_R$) unless the temperature never reached $m_\phi$.

In general, any renormalizable model that aims to \emph{explain} why the Dirac neutrino masses are so small does so by introducing new mediator particles. The couplings of these mediators to $\nu_R$ and $\nu_L$ are not tiny by construction, so will thermalize the $\nu_R$ if the temperature ever reached the mass of the mediators. Generically we then expect a contribution to  $\Delta N_\text{eff}$ in any model that addresses $m_\nu \ll m_W$.

\subsection{Leptogenesis}

Above we have argued that Dirac neutrinos could have additional interactions based on rather theoretical motivations such as Dirac stability and the smallness of neutrino masses. There is however a more pressing issue that any model of Dirac neutrinos needs to address: the baryon asymmetry of our Universe.  For \emph{Majorana} neutrinos there exist a variety of leptogenesis scenarios, in which CP-violating, out-of-equilibrium processes with $\Delta L =2$ generate a lepton asymmetry that is then transferred to a baryon asymmetry via $\Delta (B+L)=6$ sphaleron processes. For \emph{Dirac} neutrinos, there exist essentially two variations of leptogenesis:
\begin{itemize}
\item Neutrinogenesis~\cite{Dick:1999je,Murayama:2002je}: without ever breaking $B-L$, we let a new particle $\sigma$ decay out-of-equilibrium into $\nu_R$ and left-handed leptons in such a way that a lepton asymmetry is generated in the $\nu_R$ that is exactly opposite to an asymmetry in the left-handed leptons: $\Delta_{\nu_R} = - \Delta_{L}\neq 0$. If the $\nu_R$ are not thermalized afterwards a baryon asymmetry is generated out of $\Delta_L$ by the sphalerons.
\item Dirac leptogenesis~\cite{Heeck:2013vha}: breaking $B-L$ by any unit $n$ other than two makes it possible to protect the Dirac nature of neutrinos but still create a lepton asymmetry via $\Delta (B-L)=n$ interactions in complete analogy to Majorana leptogenesis mechanisms. In the simplest example with $n=4$ one creates a lepton asymmetry in $\nu_R$ via CP-violating, out-of-equilibrium decays of a new particle $\psi \to \nu_R\nu_R, \bar{\nu}_R\bar{\nu}_R$. This asymmetry in $\nu_R$ now \emph{needs} to be transferred to the left-handed leptons, e.g.~via new Yukawa interactions $\bar{L}\phi\nu_R$, in order to be further processed by the sphalerons.
\end{itemize}
From the above it is clear that Dirac leptogenesis~\cite{Heeck:2013vha} in its simplest form strongly \emph{requires} thermalized $\nu_R$, e.g.~via a neutrinophilic 2HDM, and thus predicts $\Delta N_\text{eff}\geq  0.14$. 
Neutrinogenesis~\cite{Dick:1999je} on the other hand requires the $\nu_R$ to be out-of-equilibrium \emph{after} the asymmetry generation, but unavoidably has them thermalized \emph{before}, when the mother particle $\sigma$ was still in equilibrium. For example, in the simplest realization of neutrinogenesis $\sigma$ is an electroweak doublet~\cite{Dick:1999je} and will therefore easily reach equilibrium. 
Here, too, we thus expect a contribution to $N_\text{eff}$.

In general we thus expect a $\nu_R$ contribution to $N_\text{eff}$ from any leptogenesis mechanism with Dirac neutrinos. The usual mechanisms used to evade this contribution---additional entropy dilution or a temperature below the mediator particle---would also render leptogenesis more inefficient.
Therefore, if CMB-S4 does not observe a $\Delta N_\text{eff}$ it is probably necessary to consider baryogenesis mechanisms that do not involve leptons.

\section{Conclusion}
\label{sec:conclusion}

Measurements of the radiation density in the early Universe, usually parametrized via the effective number of neutrino species $N_\text{eff}= 3.045 + \Delta N_\text{eff}$, have reached an astonishing precision within the last decade or so, thanks to experiments such as Planck. The ongoing SPT-3G experiment and the future Simons Observatory and CMB-S4 experiment will further increase our knowledge and reach sensitivities down to $\Delta N_\text{eff} \simeq 0.06$ (95\%~C.L.). This makes it possible to detect or exclude new ultralight particles even if they decoupled very early in the Universe. Here we argued that one of the best motivations for such light particles comes from the observation of neutrino oscillations. Indeed, if neutrinos are Dirac particles just like all other known fermions, we have to extend the Standard Model by two or three practically massless chirality partners $\nu_R$. Models that aim to address the Dirac stability, the smallness of neutrino masses, or the matter--anti-matter asymmetry of our Universe typically endow the $\nu_R$ with additional interactions that could lead to a thermalization in the early Universe and hence a measurable contribution to $N_\text{eff}$. The non-observation of any $\Delta N_\text{eff}$ in upcoming experiments will therefore place strong constraints on Dirac-neutrino models, as  illustrated here in some concrete examples.
On a more optimistic note, it is entirely possible that Dirac neutrinos will make themselves known in CMB $N_\text{eff}$ measurements long before their nature is confirmed in more direct ways.

\section*{Acknowledgements}
We would like to thank Michael Ratz for useful discussions.
KA and JH are supported, in part, by the National Science Foundation under Grants No.~PHY-1620638 and No.~PHY-1915005. JH is also supported in part by a Feodor Lynen Research Fellowship of the Alexander von Humboldt Foundation.

\bibliographystyle{utcaps_mod}
\bibliography{BIB}

\end{document}